\documentclass[letterpaper,aps,prl,twocolumn,groupedaddress,superscriptaddress,showpacs]{revtex4}

\usepackage{graphicx}

\begin{document}


\newcommand{\absolutefreq}{1 121 015 393 207 851(8) Hz}
\newcommand{\dnupi}{-82884(5) Hz/mT }
\newcommand{\dnusigma}{-27547(5) Hz/mT}
\newcommand{\gfS}{-0.00079248(14)}
\newcommand{\gfP}{-0.00197686(21)}
\newcommand{\gfDelta}{-0.00118437(8) }
\newcommand{\lifetimeNoError}{$\tau=20.6$ s}
\newcommand{\lifetime}{$20.6 \pm 1.4$ s}
\newcommand{\nDecays}{215 }
\newcommand{\WMIlifetime}{$22.7 \pm 4$ s}
\newcommand{\WMIgfDelta}{-0.00118(6)}

\newcommand{\AlIonShieldingFactor}{$\sigma=7.87(39)\times10^{-4}$ }
\newcommand{\AlIonBareMoment}{$\mu_N=3.64067(28)$ }
\newcommand{\AlNuclearMomentInSolution}{$\mu_N=3.64150687(65)$ }
\newcommand{\clocktransition}{$^1$S$_0$ $\rightarrow$ $^3$P$_0$ }
\newcommand{\clocktransitionm}[2]{($^1$S$_0$, m$_F$ = #1) $\rightarrow$ ($^3$P$_0$, m$_{F'}$ = #2) }
\newcommand{\clocktransitionmFhalf}[2]{($^1$S$_0$, m$_F = \frac{#1}{2}$) $\rightarrow$ ($^3$P$_0$, m$_{F'} = \frac{#2}{2}$) }
\newcommand{\clocktransitionmFhalfneg}[2]{($^1$S$_0$, m$_F = -\frac{#1}{2}$) $\rightarrow$ ($^3$P$_0$, m$_{F'} = -\frac{#2}{2}$) }
\newcommand{\gs}{$^1$S$_0$ }
\newcommand{\gsmFhalf}[1]{($^1$S$_0$, m$_{F} = \frac{#1}{2}$)}
\newcommand{\es}[1]{$^3$P$_#1$ }
\newcommand{\esns}[1]{$^3$P$_#1$}
\newcommand{\esmFhalf}[2]{$^3$P$_#1$, m$_{F'} = \frac{#2}{2}$}
\newcommand{\Al}{$^{27}$Al$^+$ }
\newcommand{\Alns}{$^{27}$Al$^+$}
\newcommand{\Be}{$^{9}$Be$^+$ }
\newcommand{\Bens}{$^{9}$Be$^+$}

\newcommand{\xfertransition}{$^1$S$_0$ $\rightarrow$ $^3$P$_1$ }
\newcommand{\xfertransitionF}{($^1$S$_0$, F $= \frac{5}{2}$) $\rightarrow$ ($^3$P$_1$, F $= \frac{7}{2}$) }
\newcommand{\xfertransitionFmF}{($^1$S$_0$, F $= \frac{5}{2}$, m$_F = \frac{5}{2}$) $\rightarrow$ ($^3$P$_1$, F $= \frac{7}{2}$,  m$_{F'} = \frac{7}{2}$) }
\newcommand{\xfertransitionPump}{($^1$S$_0$, F $= \frac{5}{2}$, m$_F = \frac{3}{2}$) $\rightarrow$ ($^3$P$_1$, F $= \frac{7}{2}$,  m$_{F'} = \frac{5}{2}$) }
\newcommand{\Betransition}{($^2$S$_{1/2}$, F = 2, m$_F = -2$) $\rightarrow$ ($^2$S$_{1/2}$, F = 1, m$_F = -1$) }
\newcommand{\Bedetection}{($^2$S$_{1/2}$, F = 2, m$_F = -2$) $\rightarrow$ ($^2$P$_{3/2}$, F = 3, m$_F = -3$) }


\title{Observation of the \clocktransition clock transition in \Al}


\author{T. Rosenband}
\email[]{trosen@boulder.nist.gov}
\author{P. O. Schmidt}
\thanks{Supported by the Alexander von Humboldt Foundation}
\thanks{Present address: Institut f\"ur Experimentalphysik, Universit\"at Innsbruck, Austria}
\author{D. B. Hume}
\author{W. M. Itano}
\affiliation{National Institute of Standards and Technology, 325
Broadway, Boulder, CO 80305}
\author{T. M. Fortier}
\affiliation{Los Alamos National Laboratory, P-23 Physics Division,
Los Alamos, NM 87545}
\author{J. E. Stalnaker}
\author{K. Kim}
\thanks{Present address: School of Mechanical Engineering, Yonsei University, Seoul, Korea}
\author{S. A. Diddams}
\author{J. C. J. Koelemeij}
\thanks{Supported by the Netherlands Organisation for Scientific Research (NWO)}
\thanks{Present address: Institut f\"ur Experimentalphysik, Heinrich-Heine-Universit\"at D\"usseldorf, Germany}
\author{J. C. Bergquist}
\author{D. J. Wineland}
\affiliation{National Institute of Standards and Technology, 325
Broadway, Boulder, CO 80305}


\date{\today}

\begin{abstract}
We report for the first time, laser spectroscopy of the
\clocktransition clock transition in \Alns.  A single aluminum ion
and a single beryllium ion are simultaneously confined in a linear
Paul trap, coupled by their mutual Coulomb repulsion. This coupling
allows the beryllium ion to sympathetically cool the aluminum ion,
and also enables transfer of the aluminum's electronic state to the
beryllium's hyperfine state, which can be measured with high
fidelity.  These techniques are applied to a measurement of the
clock transition frequency, $\nu =$ \absolutefreq.  They are also
used to measure the lifetime of the metastable clock state, $\tau =
$ \lifetime, the ground state $^1$S$_0$ g-factor, $g_S = \gfS$, and
the excited state $^3$P$_0$ g-factor, $g_P = \gfP$, in units of the
Bohr magneton.
\end{abstract}


\maketitle



The \clocktransition transition in Al$^+$ has long been recognized
as a good clock transition \cite{HGD1982monoion, Dehmelt1992}, due
to its narrow natural linewidth (8 mHz), and its insensitivity to
magnetic fields and electric field gradients (J = 0), which are
common in ion traps. More recently, this transition was also found
to have a small room-temperature blackbody radiation shift
\cite{TR2006BBRshift}. However, difficulties with direct laser
cooling and state detection have so far prevented its use.

Some of the basic ingredients for ion-trap based quantum computing
\cite{CiracZoller1995} can be used to overcome these difficulties
with the method of quantum logic spectroscopy \cite{DJW2002clocks},
where an auxiliary ion takes over the requirements of laser cooling
and state detection. Quantum logic spectroscopy was first
demonstrated experimentally on the \xfertransition transition in \Al
\cite{POS2005BeAl}.  Here we use this technique for spectroscopy of
the narrow \clocktransition clock transition, allowing, for the
first time, high precision optical spectroscopy of an atomic species
that can not be directly laser cooled.

With the advent of octave-spanning Titanium:Sapphire femtosecond
laser frequency combs \cite{SAD2000Comb, Udem2002Comb}, optical
atomic frequency standards can be compared with an uncertainty
limited only by quantum projection noise
\cite{WMI1993ProjectionNoise} and the systematic errors of the
atomic standards. This stability and accuracy can be transferred to
any part of the optical spectrum, as well as the RF domain.  The
\clocktransition clock transition in \Alns, due to its insensitivity
to external fields, is a viable candidate to reach $10^{-18}$
inaccuracy. Here we demonstrate high precision spectroscopy of the
clock transition in a single \Al ion, which is the fundamental step
necessary to realize such a frequency standard.

We use \Alns, because this is the only naturally occurring isotope
of aluminum. With nuclear spin I = $\frac{5}{2}$, \Al has no
first-order magnetic-field-independent transition, but this is not a
significant drawback. Instead, we create a ``virtual''
\clocktransitionm{0}{0} field-independent transition, by regularly
alternating between (m$_F = -\frac{5}{2})\rightarrow$(m$_{F'} =
-\frac{5}{2}$) and (m$_F = \frac{5}{2})\rightarrow$(m$_{F'} =
\frac{5}{2}$) transitions \cite{Madej1998ZeemanAverage}.
 This way, the average Zeeman state of both the
ground and excited clock state is strictly zero.  A benefit of this
approach is that the outer m$_F = \pm\frac{5}{2}$ states can be
prepared more easily than the inner m$_F$ states, by optical pumping
through the (\esns{1}, F$ = \frac{7}{2}$) state (300 $\mu$s
lifetime). In addition, the ion's Zeeman splitting serves as a
real-time magnetometer to determine the quadratic Zeeman shift of
the clock transition.

In the experiment, single \Bens, and \Al ions are loaded into a
linear Paul trap \cite{Rowe2002}, by electron impact ionization.
Under ultra-high vacuum conditions, one ion pair usually lasts for
several hours, before an adverse chemical reaction with background
gas removes one of the ions from the trap. The pair forms a two-ion
``crystal'' along the trap axis, whose in-phase-motion axial normal
mode frequency is 2.62 MHz. The radial modes (perpendicular to the
trap axis), where the \Al ion's amplitude is largest, have
frequencies of 3.8 MHz and 4.9 MHz. Laser beams of 313 nm wavelength
drive Doppler cooling, stimulated Raman, and repumping transitions
on \Be \cite{CM1995GScooling}. Another laser produces 266.9 nm
radiation to drive the \xfertransitionF transition in \Alns, and a
frequency-quadrupled fiber laser at 267.4 nm (the clock laser), with
approximately 3 Hz linewidth, excites the \clocktransition
transition.

The \clocktransition transition is probed with single interrogation
pulses from the clock laser.  Since the upper state is metastable,
the method of electron-shelving detection \cite{HGD1982monoion} is
applied. This way the \xfertransition transition is modulated by the
clock states, and the modulation condition (allowed or forbidden) is
detected with quantum logic.

\begin{figure}
\includegraphics[width=3.6in]{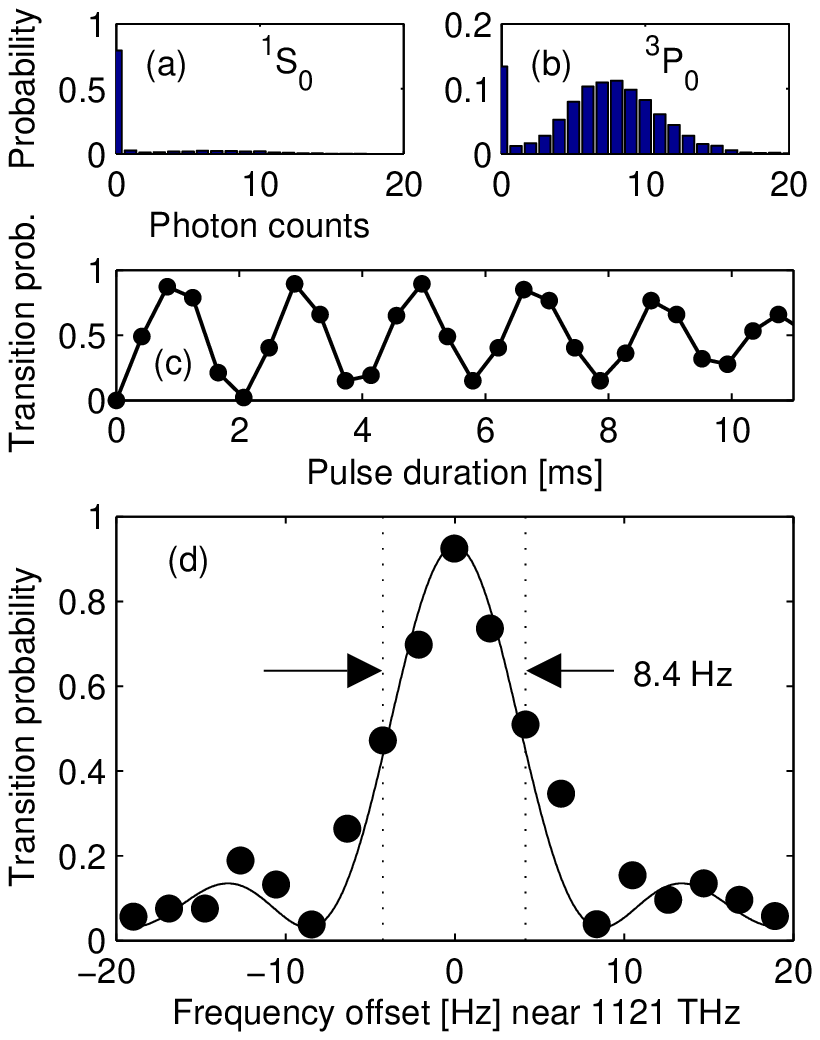}
\caption{Histograms of photon counts used for discrimination of
\gs(a) and \es{0}(b) states in Al$^+$ via Be$^+$ fluorescence (see
text).  (c) Rabi flopping averaged over forty scans of the
\clocktransition probe time (0-11 ms).   Lines are drawn to guide
the eye. (d) Average transition probability of fifty frequency scans
across the \clocktransitionmFhalf{5}{5} resonance in Al$^+$. Each
measurement consists of one $^3$P$_0$ interrogation (100 ms probe
time), followed by 5 to 10 state-detection repetitions (10 to 20 ms
total). The fitted curve is a Fourier-limited Rabi lineshape, scaled
to the observed 90 \% contrast.} \label{detection}
\end{figure}

Quantum logic spectroscopy of the \xfertransition transition has
been described previously \cite{POS2005BeAl}, and is used similarly
here.  A quantum logic pulse sequence (described below) maps the \Al
\gs state to the dark \Be  F = 1 hyperfine ground state via a
\xfertransition motional-sideband excitation. This sequence is
blocked when the \Al ion is in the \es{0} state, leaving \Be in the
bright F = 2 hyperfine ground state, into which it had previously
been optically pumped. An average of seven \Be 313 nm fluorescence
photons are counted if the \Al ion is in the \es{0} state (Figure
\ref{detection}b), and only one photon is counted if the \Al ion is
in the \gs state (Figure \ref{detection}a). This provides for
clock-state discrimination with 80 \% fidelity in a single detection
experiment, limited by inaccuracies in the various $\pi$-pulses, and
imperfect ground-state cooling. However, the single-experiment
detection fidelity does not significantly affect the final detection
fidelity for the clock state. We simply repeat the readout sequence
several times, and the combined information allows nearly unit
detection fidelity \cite{ToBePublished}.


A typical \clocktransitionmFhalf{5}{5} interrogation consists of the
following steps, with specified durations.  For m$_F$ = m$_{F'} =
-\frac{5}{2}$, the polarizations and angular momentum states of
Al$^+$ are reversed.

\begin{enumerate}
\item Sympathetic Doppler cooling via \Be (600 $\mu$s) cools all six normal
modes of the two-ion crystal to the Doppler limit (mean motional
quantum number $\bar{n} \approx 3$).
\item Clock interrogation (1 to 100 ms pulse duration, adjusted for desired resolution) drives the \clocktransition
transition.  Beryllium Doppler cooling light is applied
simultaneously to counteract anomalous heating of the ions
\cite{QAT2000heating, Deslauriers2006Heating}.
\end{enumerate}
Subsequently, state detection (\gs or \es{0}) is performed by
repeating the following sequence about 10 times. 
\begin{enumerate}
\item Optical pumping $\pi$-pulse \Al \xfertransitionPump (4 $\mu$s)
transfers residual inner Zeeman state population to the outer m$_F =
\frac{5}{2}$ state.  The inner states may be populated due to
spontaneous decay from the \es{0} state, or due to imperfect
polarization and off-resonant transitions during other
\xfertransition pulses.  Most of the time this pulse has no effect,
because \Al is already in the \gsmFhalf{5} state.
\item Sympathetic Doppler cooling (600 $\mu$s).
\item Ground-state cooling of the axial modes of the two-ion crystal
(1 ms), $\bar{n} < 0.05$ \cite{barrett03}.
\item BSB $\pi$-pulse \Al \xfertransitionFmF (30 $\mu$s).
\item RSB $\pi$-pulse \Be \Betransition (7 $\mu$s).
\item Detection on the Be$^+$ cycling transition (200 $\mu$s)
\Bedetection.  In the F = 2 state, Be$^+$ fluoresces strongly, and
it is dark in the F = 1 state (see Fig. \ref{detection}a and
\ref{detection}b).
\end{enumerate}

Because the duration of each state detection repetition is about 2
ms, excitations into the \es{1} state (300 $\mu$s lifetime) usually
decay back to the \gs state, allowing re-excitation, while \es{0}
excitations (20.6 s lifetime) are unlikely to change during the
detection sequence.

For every clock interrogation pulse, the sequence of photon counts
in step 6 is recorded by a computer, which performs a
maximum-likelihood analysis to determine whether the Al$^+$ ion was
in the \gs or \es{0} state \cite{ToBePublished}. Transitions between
the two states are detected, and multiple repetitions of the same
experiment yield the transition probability. It should be noted that
the ion may be in the \gs or \es{0} state at the beginning of each
experiment, depending on the outcome of the previous experiment
\cite{Dehmelt1990BaUpDown}. The final signal is the computer's
determination of whether or not the ion has made a transition.
Figures \ref{detection}c and \ref{detection}d show Rabi flopping,
and a scan of the clock laser frequency across the \clocktransition
resonance, with this method.

When performing an absolute frequency measurement of the
\clocktransition transition, the linear Zeeman splitting of several
$10^4$ Hz/mT (see below) in both the ground and excited clock states
must be accounted for.  To achieve first-order field insensitivity,
we create a virtual m$_F = 0$ transition by alternating between
states of opposite angular momentum, and tracking separately the
mean and difference frequencies of the two transitions
\cite{Madej1998ZeemanAverage}. The mean frequency has no linear
dependence on the magnetic field, although a quadratic Zeeman shift
of -70 Hz/mT$^2$ remains \cite{WMI2006AlCalcs}.  The difference
frequency is directly proportional to the magnetic field, providing
a real-time field measurement.  In this way, the linear Zeeman
effect does not shift the clock's center frequency, while the
quadratic Zeeman shift can be determined from the linear splitting
and accounted for with submillihertz inaccuracy.

To lock the laser frequency to the atomic reference, the laser's
frequency offset from the atomic resonance is regularly measured and
corrected. Frequency corrections are obtained by alternating between
the left and right ``half-power'' points of the ion's resonance and
applying frequency feedback to keep the transition probabilities
equal.
 In order to switch between angular momentum states, a computer-controlled waveplate selects purely
$\sigma_+$ or $\sigma_-$-polarized \xfertransition light that
optically pumps \Al to the m$_F = \pm\frac{5}{2}$ ground states. At
the operating field of approximately 0.1 mT, and typical Fourier
limited linewidths of 20 Hz or less, the linear Zeeman structure of
both clock states is well resolved. A synthesizer driven
acousto-optic frequency shifter changes the probe frequency by
$\Delta f\approx -4$ kHz to drive the \clocktransitionmFhalf{5}{5}
transition, and by $-\Delta f$ to drive the opposite Zeeman states.
When locking the clock laser to the atomic transition, the clock
ground state is switched between the m$_F = \pm\frac{5}{2}$ states
every three seconds. This way, magnetic field fluctuations
(typically below 100 nT in several minutes) contribute only
short-term instability in the clock frequency, rather than long-term
inaccuracy.

We have locked the clock laser to the virtual m$_F$=0 clock
transition as described above and used the fourth subharmonic at
1070 nm to reference one tooth of an octave-spanning
Titanium:Sapphire femtosecond laser frequency comb
(femtosecond-comb) \cite{Fortier2006BBcomb}. The offset frequency of
the femtosecond-comb is locked by use of an f-2f interferometer, and
the repetition rate is measured by a hydrogen maser referenced to
the NIST-F1 primary cesium standard \cite{Jefferts2005F1}.  A 2600
second long measurement of the \Al clock transition frequency yields
$\nu $ = \absolutefreq, where the uncertainty is dominated by
short-term frequency fluctuations in the hydrogen maser.  Systematic
uncertainties in Al$^+$ are not expected to be significant at this
level, where the largest terms are second-order Doppler shifts of
approximately 0.03 Hz.

\begin{figure}
\includegraphics[width=3.4in]{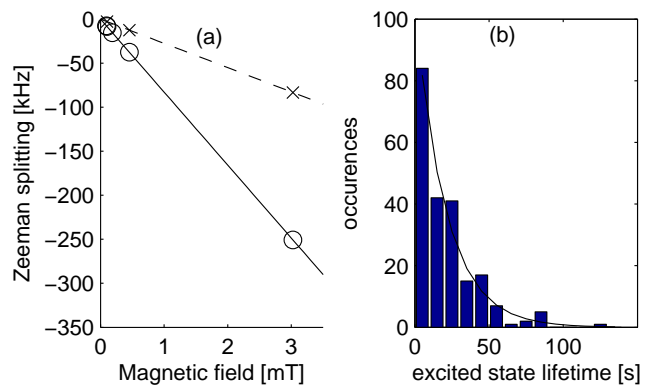}
\caption{Linear Zeeman splitting (a) of the $(m_F =
\pm\frac{5}{2})\rightarrow(m_F' = \pm\frac{5}{2})$ transition pair
(o) and the $(m_F = \pm\frac{5}{2})\rightarrow(m_F' =
\pm\frac{3}{2})$ transition pair (x). Straight lines correspond to
the reported g-factor differences. Histogram of lifetimes (b) of
\nDecays decays from the $^3$P$_0$ state.  The curve shows the
expected number of decays for a lifetime \lifetimeNoError.}
\label{Decay}
\end{figure}

We have also measured the Land\'{e} g-factors of the \gs and \es{0}
states. Interrogation pulses ($\pi$-polarized) yield the g-factor
difference $g_P - g_S$ \cite{Boyd2006Coherence}, where $g_P$ and
$g_S$ are the g-factors of the \es{0} and \gs states, respectively.
In addition we use $\sigma_{+/-}$-polarized interrogation pulses to
measure another, linearly independent combination of $g_P$ and
$g_S$, providing an accurate measurement of both g-factors, which is
unaffected by the chemical shifts that currently limit the accuracy
of nuclear magnetic resonance (NMR) based measurements of $g_S$
\cite{Gustavsson1998NMR}. An external magnetic field B splits the
m$_F$=$\pm\frac{5}{2}$ transition frequencies by $\Delta\nu = 5 B
(g_P - g_S)\mu_B$, where $\mu_B$ is the Bohr magneton. The frequency
splitting $\Delta\nu$ is tracked and recorded automatically in the
clock laser lock described above. In order to measure the magnetic
field, we use the simultaneously trapped Be$^+$ ion.  Be$^+$ has
well known ground-state Zeeman splitting \cite{DJW1983BeHFS},
allowing for accurate calibration of the external magnetic field. In
between aluminum clock interrogations, a 1.25 GHz magnetic field is
applied to the beryllium ion with a loop antenna. Computer control
software adjusts the radiation frequency to drive and track the
magnetic-field-sensitive \Betransition transition, establishing a
record of the magnetic field.  We find the Zeeman splitting of \Al
to be $\Delta\nu/B$ = \dnupi, or $g_P - g_S$ = \gfDelta (see Figure
\ref{Decay}a).  A recent multi-configuration Dirac-Hartree-Fock
calculation \cite{WMI2006AlCalcs} yields a similar theoretical value
of $g_P - g_S$ = \WMIgfDelta.

In order to determine separate values $g_P$, and $g_S$, we  repeated
the above experiment with $\sigma_{+/-}$-polarized clock radiation.
This way, the \clocktransitionmFhalfneg{5}{3}, and
\clocktransitionmFhalf{5}{3} transitions are probed, and their
magnetic-field-induced splitting $\Delta\nu'$ is measured. From
$\Delta\nu'/B = (3g_P - 5g_S)\mu_B =$ \dnusigma $ $ we find $g_P =
\gfP$ and $g_S = \gfS$.  The uncertainties of these values are
dominated by inaccuracies in the \Bens-based magnetic field
calibration.  This was caused by a perturbation of the current
supply for the DC magnetic field from the 1.25 GHz RF pulse itself.
We estimate that this perturbation is $21\pm200$ nT in both the
$\pi$ and $\sigma_{+/-}$ measurements, and derive the stated
uncertainties from the measurements at 3 mT, where the magnetic
field uncertainty has the smallest effect on our results.

Since $g_S$ depends solely on the ion's shielded nuclear magnetic
moment, this measurement may be compared with the value recorded in
nuclear data tables, which stems from NMR experiments.  An estimated
shielding factor of \AlIonShieldingFactor \cite{WMI2006AlCalcs} for
\Alns, leads to a bare moment of \AlIonBareMoment nuclear magnetons.
This differs from the published value of \AlNuclearMomentInSolution
\cite{Raghavan1989} by 0.00084(28).  However, the liquid-state NMR
result does not account for the chemical shift of the nuclear
magnetic moment due to the aqueous solution, which generally has a
magnitude of $10^{-4}-10^{-3}$ nuclear magnetons
\cite{Gustavsson1998NMR}.

Hyperfine-induced spontaneous decay rates of \es{0} states for the
Mg-isoelectronic sequence have been calculated
\cite{Brage1998Al3P0lifetime}.  More recently, the lifetime was
calculated \cite{WMI2006AlCalcs} as \WMIlifetime. Here we measured
the lifetime by repeatedly exciting the \es{0} state and waiting for
spontaneous decay into the ground state \cite{Peik1994In}. The
aluminum internal state was monitored by continuously applying the
previously described clock state readout sequence. Once decay has
been detected, the ion is re-excited.  A total of \nDecays
spontaneous emission events were observed this way, with a mean
lifetime $\tau$ = \lifetime. The reported uncertainty is
statistical, and systematic biases are estimated to be less than
0.25 s. Figure \ref{Decay}(b) shows the distribution of decay times.

In conclusion, we have demonstrated high resolution spectroscopy of
the \clocktransition clock transition in \Al using quantum logic.
This technique was used to measure the clock transition frequency
with a fractional uncertainty of $7\times10^{-15}$.  We have also
calibrated the g-factors of the ground and excited states and
measured the excited state lifetime.  These are the basic steps
necessary to implement an accurate frequency standard based on
\Alns.

\begin{acknowledgments}
We acknowledge support from ONR and DTO. We thank V. Meyer for
constructing the ion trap used for these experiments, and E. A.
Donley, J. Ye, M. A. Lombardi, and D. R. Smith for their careful
reading of this manuscript. The authors are grateful to T. Parker,
S. Jefferts and T. Heavner for providing the absolute frequency
calibration. This work is a contribution of NIST, and is not subject
to U.S. copyright.
\end{acknowledgments}


\end{document}